\documentclass[prb,letterpaper,aps,floatfix,twocolumn]{revtex4-1}
\usepackage{graphicx}
\usepackage{amsmath}
\usepackage{amsfonts}
\usepackage[small,bf]{subfigure}
\newcommand{\beq}{\begin{equation}}
\newcommand{\eeq}{\end{equation}}

\newcommand{\beqa}{\begin{eqnarray}}
\newcommand{\eeqa}{\end{eqnarray}}
\newcommand{\pdg}{{\vphantom \dag}}
\newcommand{\dg}{{\dag}}

\newcommand{\upa}{\uparrow}
\newcommand{\da}{\downarrow}

\newcommand{\cL}{{\cal L}}
\begin{document}
\title{Theory of the fractional quantum Hall effect in Weyl semimetals}
\author{Manisha Thakurathi}
\affiliation{Department of Physics and Astronomy, University of Waterloo, Waterloo, Ontario 
N2L 3G1, Canada} 
\author{A.A. Burkov}
\affiliation{Department of Physics and Astronomy, University of Waterloo, Waterloo, Ontario 
N2L 3G1, Canada} 
\date{\today}
\begin{abstract}
We develop a hydrodynamic field theory of the three-dimensional fractional
quantum Hall effect, which was recently proposed to exist in magnetic Weyl semimetals, when the Weyl nodes 
are gapped by strong repulsive interactions. This theory takes the form of a BF theory, which contains both one-form and two-form gauge fields, 
coupling to quasiparticle and loop excitations correspondingly. It may be regarded as a generalization of the Chern-Simons theory of two-dimensional fractional quantum Hall liquids to 
three dimensions. 
\end{abstract}
\maketitle
\section{Introduction}
\label{sec:1}
One of the fundamental properties of any quantum mechanical system is level repulsion: energy levels do not cross 
when parameters are varied. An important implication of this is that gapless excitations in many-body systems are 
an exception rather than a rule, in the sense that they do not appear without a specific reason.
In particular, conventional phases of matter are distinguished by spontaneously broken symmetries. These necessarily 
lead to gapless excitations, or Goldstone modes, described by the corresponding ``nonlinear sigma model". 
When the Goldstone modes fluctuate strongly and the broken symmetry is restored, a gap is generally opened.~\cite{Haldane83}

Gapless spectrum may also arise for topological, rather than symmetry, reasons.~\cite{Volovik03,Volovik07}
Topological insulators (TI) necessarily have gapless states on their boundaries,~\cite{Haldane88,Hasan10,Qi11}
which are required by the existence of nontrivial invariants, characterizing the bulk electronic structure of the insulator. 
But topology-mandated gaplessness is not limited to TI surfaces. Bulk spectrum may also be gapless for topological reasons.
The best known and simplest example of this is the fact that when the number of electrons per unit cell of a crystal 
is an odd integer, the material must be a metal with a Fermi surface of gapless particle-hole excitations, whose 
enclosed volume in momentum space is proportional to the electron density.~\cite{Luttinger60}
The only way to avoid gaplessness in this situation, without breaking translational symmetry and thus changing the number of electrons per unit cell, is through the formation of a Mott insulator, in which electron quantum numbers are fractionalized and topological order, i.e. genus-dependent ground state degeneracy, is present.~\cite{Wen90,LSM,Oshikawa00,Hastings04}

When the number of electrons per unit cell is even, conventional wisdom holds that one may (and often does) still get a metal due to accidental band overlap, but 
generically one gets an insulator, as the Luttinger volume vanishes in this case. 
This conventional wisdom was shown to be incorrect recently, when Weyl semimetals were discovered.~\cite{Weyl_RMP,Burkov_ARCMP,Felser_ARCMP,Hasan_ARCMP}
A Weyl semimetal has a gapless bulk spectrum when the number of electrons per unit cell is even and one thus normally expects an insulator, or an accidental 
(semi)metal with zero Luttinger volume, which may be deformed into an insulator by a small perturbation of the Hamiltonian. 
In contrast, gapless spectrum in a Weyl semimetal is mandated by topology and the Weyl semimetal phase arises unavoidably in certain generic situations, in particular as 
an intermediate phase between TI and normal insulator in three dimensions (3D) when either time reversal (TR) or inversion symmetries are violated.~\cite{Murakami07,Burkov11-1}

In analogy to the Mott insulator state in strongly correlated materials with an odd number of electrons per unit cell, one may ask whether we can circumvent the topologically-mandated gaplessness in Weyl semimetals and open a gap when strong electron-electron interactions are introduced. Is this possible and, if yes, what is the nature of the insulating state one obtains? We asked and answered this question in Ref.~\onlinecite{Wang20} in the context of the simplest realization of a Weyl semimetal with a single pair of opposite-chirality 
Weyl nodes at the Fermi energy (see Refs.~\onlinecite{Meng16,Morimoto16,Sagi18,Meng19,Teo19} for alternative discussions of this problem). 
Such a Weyl semimetal inevitably arises as an intermediate phase between an integer quantum Hall (IQH) and normal insulator in 3D. 

In a 3D band insulator the Hall conductivity is quantized as
\beq
\label{eq:1}
\sigma_{xy} = \frac{e^2}{h} \frac{G}{2\pi}, 
\eeq
where $G$ is a reciprocal lattice vector.~\cite{Halperin87,Kohmoto92,Volovik19}
Just as in 2D, this integer (i.e. integer multiple of a primitive reciprocal lattice vector $2 \pi/a$, where $a$ is the lattice constant)
quantization is a direct and inevitable consequence of gauge invariance. 
Suppose we want to realize a transition between a 3D normal insulator with $\sigma_{xy} = 0$ and an IQH insulator with $\sigma_{xy} = e^2/h a$. 
In 2D the analogous transition is a sharp ``plateau transition": the Hall conductivity jumps between the two quantized values. 
This sharp transition is a consequence of the fact that the Hall conductivity in 2D is dimensionless in units of a combination of fundamental constants $e^2/h$ and there is no 
way to smoothly interpolate between the two quantized values. 
In 3D, however, the situation is different and the Hall conductivity involves a wavevector. This implies that the IQH transition does not have to be (in fact can not be) sharp in 3D. 
Instead it proceeds through a gapless phase, a Weyl semimetal, in which the Hall conductivity is given by
\beq
\label{eq:2}
\sigma_{xy} = \frac{e^2}{h} \frac{2 Q}{2 \pi}, 
\eeq
where $2 Q$ is the separation between a pair of opposite-chirality Weyl nodes in momentum space, which changes smoothly between $0$ and $G$. 
Vice versa, such a ``fractional" Hall conductivity in the absence of a Fermi surface (Luttinger volume is zero due to even number of electrons per unit cell)
inevitably, by gauge invariance, implies the presence of gapless Weyl nodes. 

Another very useful viewpoint on the connection between noninteger Hall conductivity in 3D and Weyl nodes is provided by the concept of the 
chiral anomaly.~\cite{Adler69,Jackiw69,Nielsen83,Zyuzin12-1}
In the absence of the Fermi surface, the Hall conductivity is a thermal equilibrium property, given by the derivative of the electron density with respect to the magnetic field
\beq
\label{eq:3}
\sigma_{xy} = \left(\frac{\partial n}{\partial B} \right)_{\mu},
\eeq
where we will switch to $\hbar = c = e = a =1$ units henceforth. 
The nonzero derivative with respect to the applied magnetic field arises in the Weyl semimetal case due to the property that the lowest Landau level 
crosses the Fermi energy at the locations of the two Weyl points (the classic chiral anomaly, i.e. nonconservation of the chiral charge, is a direct consequence of that). 
This introduces an effective 1D metal with a magnetic-field-dependent electron density
\beq
\label{eq:4}
n = \frac{2 Q}{2 \pi} \frac{B}{2 \pi}, 
\eeq
which is not an integer per $2 \pi \ell_B^2 = 2 \pi/ B$, leading to a nonzero 1D Luttinger volume. 
The derivative of this extra magnetic-field-induced Luttinger volume with respect to the magnetic field gives 
the ``fractional" Hall conductivity of Eq.~\eqref{eq:2}, thus revealing a connection between gaplessness of Weyl semimetals, 
chiral anomaly and the Luttinger's theorem.~\cite{Song19}

Using ``vortex condensation" method,~\cite{Wang13,Metlitski15} we found in Ref.~\onlinecite{Wang20} that Weyl nodes may indeed be gapped out while preserving 
the chiral anomaly, i.e. the electrical and thermal Hall conductivities of the gapless Weyl semimetal, when the separation between the Weyl nodes is exactly half the 
reciprocal lattice vector $2Q = \pi$. 
The resulting state was shown to be a 3D version of a nonabelian fractional quantum Hall state, which may be viewed as a 3D TR-breaking analog of the Pfaffian-antisemion 
state on the surface of a 3D TI.~\cite{Wang13,Metlitski15}
In this paper we provide mathematical details of the vortex condensation procedure and derive a hydrodynamic BF theory~\cite{Zee94,Franz07,Cho11,Zaanen11,Vishwanath13,Ryu14,Ye15} 
of this novel 3D FQH state. 
This theory may be viewed as a 3D analog of the Chern-Simons field theory of 2D FQHE. 

\section{Derivation of the hydrodynamic theory}
\label{sec:2}
To derive the hydrodynamic theory of the 3D FQH liquid microscopically, we start from the simplest lattice model of a magnetic Weyl semimetal with two nodes.~\cite{Burkov11-1,Trivedi17}
The momentum space Hamiltonian is given by
\beq
\label{eq:5}
H = \sum_k \psi^\dg_k \left[\sigma_x \sin(k_x) + \sigma_y \sin(k_y) + \sigma_z m(k)\right] \psi^\pdg_k. 
\eeq
Here $\sigma_i$ are Pauli matrices, describing the pair of touching bands and
\beq
\label{eq:6}
m(k) = \cos(k_z) - \cos(Q) - \tilde m [2 - \cos(k_x) - \cos(k_y)], 
\eeq
where $\tilde m > 1$ and $m(k)$ vanishes at two points on the $z$-axis with $k_z = \pm Q$, which are the locations of the Weyl nodes.
Eq.~\eqref{eq:5} has the form of a Hamiltonian of a massive 2D Dirac fermion with mass $m(k)$, which changes sign 
at the Weyl node locations. 
Since a massive 2D Dirac fermion contributes $\textrm{sign} (m)/4 \pi$ to the Hall conductivity, it follows from Eqs.~\eqref{eq:5}, \eqref{eq:6} that 
the Hall conductivity of the Weyl semimetal is given by Eq.~\eqref{eq:2}. 

Fourier transforming Eq.~\eqref{eq:5} to real space and coupling to external electromagnetic field, we obtain
\beqa
\label{eq:7}
H&=&\sum_r \left\{i A_{r0} \psi^\dg_r \psi^\pdg_r - [\cos(Q) + 2 \tilde m]\psi^\dg_r \sigma_z  \psi^\pdg_r \right. \nonumber \\
&-&\left.\frac{i}{2} \psi^\dg_r \left(\sigma_x + i \tilde m \sigma_z\right) \psi^\pdg_{r + x} e^{i A_{r x}} + h.c. \right. \nonumber \\
&-& \left. \frac{i}{2} \psi^\dg_r \left(\sigma_y + i \tilde m \sigma_z\right) \psi^\pdg_{r + y} e^{i A_{r y}} + h.c. \right. \nonumber \\ 
&+&\left.\frac{1}{2} \psi^\dg_r \sigma_z  \psi^\pdg_{r + z} e^{i A_{r z}} + h.c. \right\}.
\eeqa
We now use parton representation of the electron operators~\cite{Georges_SR}
\beq
\label{eq:8}
\psi_r = e^{i \theta_r} f_r, 
\eeq
where $e^{i \theta_r}$ represents a spinless charged boson (chargon) while $f_r$ is a two-component neutral fermion (spinon) which carries the remaining spin and orbital quantum numbers of the electron. 
The spinon number satisfies a local constraint
\beq
\label{eq:9}
f^\dg_r f^\pdg_r = n_r, 
\eeq
where $n_r$ is the chargon number operator, conjugate to the phase
\beq
\label{eq:10}
[\theta_r, n_r] = -i. 
\eeq

Decoupling the spinon and chargon variables using Hubbard-Stratonovich transformation, we obtain the following contributions to the imaginary time Lagrangian 
density $\cL = \cL_f + \cL_b$ (the imaginary time action is $S = \int_0^\beta d \tau \sum_r \cL$)~\cite{Lee_SR,Senthil_SR,Barkeshli12} 
\beqa
\label{eq:11}
\cL_f&=&f^\dg_r (\partial_{\tau} - i a_{r 0}) f^\pdg_r  - [\cos(Q) + 2 \tilde m] f^\dg_r \sigma_z  f^\pdg_r \nonumber \\
&-&\frac{i \chi}{2} f^\dg_r \left(\sigma_x + i \tilde m \sigma_z\right) f^\pdg_{r + x} e^{- i a_{r x}} + h.c. \nonumber \\
&-&\frac{i \chi}{2} f^\dg_r \left(\sigma_y + i \tilde m \sigma_z\right) f^\pdg_{r + y} e^{- i a_{r y}} +h.c. \nonumber \\
&+&\frac{\chi}{2} f^\dg_r \sigma_z  f^\pdg_{r + z} e^{- i a_{r z}} + h.c.,
\eeqa
and 
\beq
\label{eq:12}
\cL_b = i n_r (\partial_{\tau} \theta_r + A_{r 0} + a_{r 0}) - \chi \cos(\Delta_i \theta_r + A_{r i} + a_{r i}). 
\eeq
Here $\chi$ and $a_{r i}$ are the amplitude and the phase of the Hubbard-Stratonovich field, coupling chargons and spinons, $a_{r 0}$ is a Lagrange multiplier which 
enforces the constraint in Eq.~\eqref{eq:9}, 
and $a_{r \mu}$ is thus a compact U$(1)$ gauge field, associated with the U$(1)$ gauge invariance implicit in the parton decomposition~\eqref{eq:8}.
$\Delta_i \theta_r = \theta_{r + i} - \theta_r$ is a discrete derivative. 
We have not explicitly included the electron-electron interaction terms, which are generally present in both $\cL_f$ and $\cL_b$. 
While the specific form of these interactions is not important for our purposes, their presence certainly is, since the physics we will discuss can arise only as a result of repulsive 
electron-electron interactons. 

In the language of the parton construction, ``vortex condensation" means pairing spinons (which by itself does not produce a superconducting state since the spinons 
are neutral) and condensing vortices of the chargon field to produce a charge insulator. This procedure results in a gapped incompressible state, if no symmetries 
are broken. 
As discussed in Ref.~\onlinecite{Wang20}, to gap out the spinons, finite-momentum (Fulde-Ferrell-Larkin-Ovchinnikov, or FFLO) pairing is necessary, where spinons on each side of the left- and right-handed Weyl cone are paired.~\cite{Moore12,Bednik15} This generally produces a state with broken translational symmetry. 
Since the pairing field carries momentum $2 Q$, a gauge-invariant density modulation carries momentum $4 Q$. 
This means that when and only when $4 Q = G = 2 \pi$, the gapped state of paired spinons actually does not break translational symmetry. 
This gives $2 Q = \pi$, i.e. the Weyl node separation of exactly half the reciprocal lattice vector. 
The Fermi arc of the spinon Weyl semimetal turns into a Majorana surface state in this case, which spans the entire Brillouin zone (BZ), allowing the Weyl nodes to be gapped without 
BZ reduction and thus translational symmetry breaking. 

Alternatively, this state, specifically in the case $2 Q = \pi$, may also be obtained using BCS zero-momentum pairing.~\cite{Meng12,Moore12,Bednik15,YiLi18} This route was not discussed in Ref.~\onlinecite{Wang20}, so let us 
elaborate on it here. 
Let us start from the spinon Hamiltonian, ignoring the coupling to chargons, and add standard BCS pairing of time-reversed states
\beqa
\label{eq:13}
H&=&\sum_k f^\dg_k \left[\sigma_x \sin(k_x) + \sigma_y \sin(k_y) + m(k) \sigma_z\right] f^\pdg_k \nonumber \\
&+& \Delta \sum_k (f^\dg_{k \upa} f^\dg_{-k \da} + f^\pdg_{-k \da} f^\pdg_{k \upa}). 
\eeqa
Introducing Nambu vector $\tilde f_k = (f^\pdg_{k \upa}, f^\pdg_{k \da}, f^\dg_{-k \da}, f^\dg_{-k \upa})$ and diagonalizing the particle-hole block of the Hamiltonian, 
we obtain 
\beqa
\label{eq:14}
H&=&\frac{1}{2} \sum_k \tilde f^\dg_k \left\{\sigma_x \sin(k_x) + \sigma_y \sin(k_y)\right. \nonumber \\
&+&\left.[m(k) \pm \Delta] \sigma_z\right\} \tilde f^\pdg_k. 
\eeqa
When $2 Q = \pi$ and $\Delta > 1$ this describes a fully gapped topological superconductor, which may be viewed as a stack of 2D $p+ip$ superconductors 
and has a Majorana mode, spanning the BZ in the $z$-direction.~\cite{Meng12,Bednik15}
This is identical to the state one gets by gapping out the Weyl nodes with FFLO pairing when $2 Q = \pi$. Henceforth we will assume that the Weyl node separation is $2 Q = \pi$ and the spinons are gapped by either FFLO or strong BCS pairing. 

\begin{figure}[t]
\includegraphics[width=8.5cm]{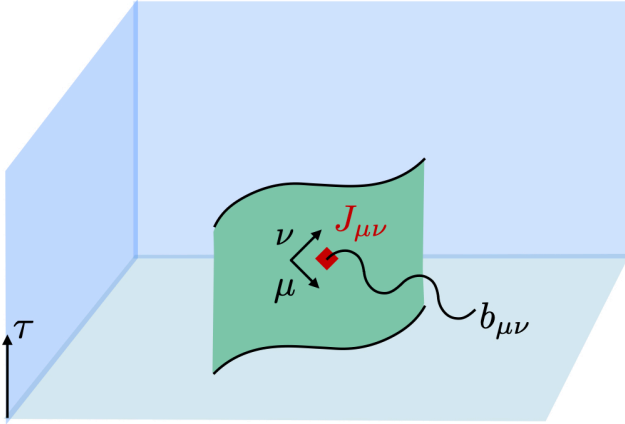}
\caption{(Color online) Cartoon of a vortex worldsheet in $3+1$ dimensions, swept by a section of a vortex loop. The two-form vortex current $J_{\mu \nu}$ is associated with a tangent plane to the worldsheet at a particular 
space-time point. It is minimally coupled to the two-form gauge field $b_{\mu \nu}$. The two-form gauge invariance is connected to the vorticity conservation.}
\label{fig:1}
\end{figure}

We now turn to the charge sector of the theory, described by $\cL_b$, which contains most of the physics of the vortex condensation. 
To describe vortex condensation we need to pass to the dual description. 
We start by decoupling the cosine in Eq.~\eqref{eq:12} using Villain transformation, which is in essence a discrete analog of the Hubbard-Stratonovich transformation
\beqa
\label{eq:15}
\cL_b&=& i n_r (\partial_{\tau} \theta_r + A_{r 0} + a_{r 0}) \nonumber \\
&+&i J_{r i} (\Delta_i \theta_r + A_{r i} + a_{r i}) + \frac{1}{2 \chi} J_{r i}^2 + \frac{1}{2 \chi} n_r^2, 
\eeqa
where the currents $J_{r i}$ are integer and the last term arises from repulsive electron-electron interactions. 
We have made its coefficient the same as that of the $J_{r i}^2$ term for notational brevity, this does not lead to any loss of generality. 
Identifying $n_r = J_{r 0}$ and discretizing the imaginary time, Eq.~\eqref{eq:15} may be written in a compact ``relativistic" notation
\beq
\label{eq:16}
\cL_b = i J_{r \mu} (\Delta_{\mu} \theta_r + A_{r \mu} + a_{r \mu}) + \frac{1}{2 \chi} J^2_{r \mu}. 
\eeq
The currents $J_{r \mu}$ are defined on the corresponding links of the space-time lattice. 
Integrating over $\theta_{r}$ produces the chargon current conservation law
\beq
\label{eq:17}
\Delta_{\mu} J_{r \mu} = 0. 
\eeq
This may be solved as 
\beq
\label{eq:18}
J_{\mu} = \frac{1}{2 \pi} \epsilon_{\mu \nu \lambda \rho} \Delta_{\nu} b_{\lambda \rho}, 
\eeq
where $b_{\mu \nu}$ is a $2 \pi \times$integer-valued antisymmetric two-form gauge field, defined on plaquettes of the dual space-time lattice, bisecting the links of the 
direct lattice on which the currents $J_{\mu}$ are defined. For brevity we will drop the $r$ subscripts henceforth. 
Eq.~\eqref{eq:18} possesses gauge invariance with respect to the transformation $b_{\mu \nu} \rightarrow b_{\mu \nu} + \Delta_{\mu} g_{\nu} - \Delta_{\nu} g_{\mu}$. 
The two-form gauge field $b_{\mu \nu}$ is minimally coupled to a conserved two-form vortex space-time current $J_{\mu \nu}$, to be explicitly introduced below, which describes vortex worldsheets in $3+1$ space-time dimensions, see Fig.~\ref{fig:1}.~\cite{Zee94,Franz07,Cho11,Zaanen11,Vishwanath13,Ryu14,Ye15}

It is convenient to relax the $2 \pi \times$integer constraint on $b_{\mu \nu}$ by introducing a ``vortex kinetic energy term", as follows~\cite{Motrunich05}
\beqa
\label{eq:19}
\cL_b&=&\frac{i}{2 \pi} (A_{\mu} + a_{\mu}) \epsilon_{\mu \nu \lambda \rho} \Delta_{\nu} b_{\lambda \rho} + \frac{1}{8 \pi^2 \chi} 
(\epsilon_{\mu \nu \lambda \rho} \Delta_{\nu} b_{\lambda \rho})^2 \nonumber \\
&-&t \cos(\Delta_{\mu} \alpha_{\nu} - \Delta_{\nu} \alpha_{\mu} + b_{\mu \nu}). 
\eeqa
This term, which is manifestly gauge invariant, describes vorticity-conserving motion of vortex loops. 
The operator $e^{- i \alpha_{r \mu}}$ creates a segment of a vortex loop on the link $(r, \mu)$ of the dual lattice [the $r$ indices are suppressed in \eqref{eq:19}]. 
The kinetic energy term then describes annihilation of vortex loop segments on links $(r, \mu)$ and $(r+ \mu, \nu)$ and simultaneous creation of two segments on links 
$(r, \nu)$ and $(r + \nu, \mu)$, which is the simplest possible process, changing the local geometry of the vortex loop, while preserving its continuity. 

Conventional vortex condensation would mean taking $t$ to be large, in which case the last term in Eq.~\eqref{eq:19} results in Meissner effect for the 
gauge field $b_{\mu \nu}$ giving rise to a mass term $b_{\mu \nu} b_{\mu \nu}$. 
This, however, produces a trivial Mott insulator with a zero electrical Hall conductivity, not the state we are looking for, preserving the chiral anomaly. 
In order to fix this, it is convenient to switch to a dual view of the vortex condensation, which operates explicitly with vortex currents. 
As before, we decouple the cosine using Villain transformation 
\beqa
\label{eq:20}
{\cal L}_b&=&\frac{i}{2 \pi} (A_{\mu} + a_{\mu}) \epsilon_{\mu \nu \lambda \rho} \Delta_{\nu} b_{\lambda \rho} + \frac{1}{8 \pi^2 \chi} 
(\epsilon_{\mu \nu \lambda \rho} \Delta_{\nu} b_{\lambda \rho})^2 \nonumber \\
&+&i J_{\mu \nu} (\Delta_{\mu} \alpha_{\nu} - \Delta_{\nu} \alpha_{\mu} + b_{\mu \nu}) + \frac{1}{2 t} (J_{\mu \nu})^2, 
\eeqa
where $J_{\mu \nu}$ are integer vortex current variables. 
Integrating out $\alpha_{\mu}$ gives the vorticity conservation law
\beq
\label{eq:21}
\Delta_{\mu} J_{\mu \nu} = 0, 
\eeq
which may be solved as
\beq
\label{eq:22}
J_{\mu \nu} = \frac{1}{2 \pi} \epsilon_{\mu \nu \lambda \rho} \Delta_{\lambda} c_{\rho}, 
\eeq
where $c_{\mu}$ is a $2 \pi \times$integer valued one-form gauge field, defined on the links of the direct space-time lattice, 
perpendicular to the plaquettes of the dual lattice on which currents $J_{\mu \nu}$ are defined. 
Then Eq.~\eqref{eq:20} becomes
\beqa
\label{eq:23}
{\cal L}_b&=&\frac{i}{2 \pi} (A_{\mu} + a_{\mu} + c_{\mu}) \epsilon_{\mu \nu \lambda \rho} \Delta_{\nu} b_{\lambda \rho} \nonumber \\
&+&\frac{1}{8 \pi^2 \chi} (\epsilon_{\mu \nu \lambda \rho} \Delta_{\nu} b_{\lambda \rho})^2 + 
\frac{1}{8 \pi^2 t} (\epsilon_{\mu \nu \lambda \rho} \Delta_{\lambda} c_{\rho})^2. 
\eeqa
The $2 \pi \times$integer constraint on $c_{\mu}$ may be softened as before by introducing a term $- \tilde t \cos(\Delta_{\mu} \phi + c_{\mu})$, where $\phi$ is 
essentially the phase of the chargons. Since we want to describe a charge insulator, the chargons are gapped and therefore this term may be ignored. 
One needs to remember, however, that $c_{\mu}$ is a compact gauge field, defined modulo $2 \pi$. 

Then, integrating over $c_{\mu}$ produces a Meissner term for $b_{\mu \nu}$ and a trivial Mott insulator. 
This would correspond to ordinary condensation of $2 \pi$ vortices. 
Instead, we will condense double (flux $4 \pi$) vortices, placing them in a quantum Hall rather than a simple superfluid state. 
To accomplish this, we replace the $2 \pi$ vortex kinetic energy term in Eq.~\eqref{eq:19} with the $4 \pi$ vortex kinetic energy term 
$-t \cos(\Delta_{\mu} \alpha_{\nu} - \Delta_{\nu} \alpha_{\mu} + 2 b_{\mu \nu})$. 
Carrying out Villain transformation and integrating out $\alpha$ as in Eqs.~\eqref{eq:20}--\eqref{eq:23}, we obtain
\beq
\label{eq:24}
{\cal L}_b = \frac{i}{2 \pi} (A_{\mu} + a_{\mu} + 2 c_{\mu}) \epsilon_{\mu \nu \lambda \rho} \partial_{\nu} b_{\lambda \rho} - \frac{2 i}{4 \pi} \epsilon_{z \mu \nu \lambda} 
c_{\mu} \partial_{\nu} c_{\lambda} + \ldots, 
\eeq
where $\ldots$ stand for the Maxwell terms and we have taken the continuum limit. 
The factor of $2$ in front of $c_{\mu}$ in the first term expresses the fact that we are condensing flux $4 \pi$ vortices. 
The first term in Eq.~\eqref{eq:24} is the standard topological BF term, which simply encodes the mutual phase factors of particles and flux $4\pi$ vortices. 
The second term, which we added by hand, is a Chern-Simons term for $c_{\mu}$. It has a properly quantized coefficient $2/ 4 \pi$, which ensures gauge 
invariance. Quantization of the coefficient may be established by viewing the contribution of the second term in Eq.~\eqref{eq:24} to the imaginary time action as a sum of 
standard 2D Chern-Simons terms over atomic planes, stacked in the $z$-direction. 
This topological term expresses the fact that the condensed double vortices exist not in a simple superfluid, but in a 3D quantum Hall state, whose physics 
we describe in detail below. 

We need to note here that a more mathematically complete coordinate-independent formulation of the topological field theory in Eq.~\eqref{eq:24} should involve 
a ``translation gauge field",~\cite{Liu13,Vozmediano15,Franz16,Grushin16,Volovik19,Wang20,Song19} describing elastic response of the 3D crystal. This is a consequence of the fact that a Weyl semimetal 
is protected by the crystal translational symmetry and we are aiming to describe a 3D featureless liquid state, obtained by gapping the Weyl nodes without violating the translational 
symmetry. This naturally leads to the emergence of elastic gauge fields, whose fluxes are related to the density of crystalline defects (dislocations). 
However here, in the interests of clarity and simplicity, we will focus on the electromagnetic (and thermal) response and thus choose a more simple-minded formulation, 
in which we explicitly fix the direction of the Weyl node separation vector to be the $z$-direction. This fixes the first index of the antisymmetric tensor in the 
Chern-Simons term in Eq.~\eqref{eq:24}. We leave the more complete formulation of the theory, involving elastic gauge fields, to future work. 

\section{Physics of the 3D fractional quantum Hall liquid}
\label{sec:3}
Let us see that this theory indeed describes the correct physics, which was introduced in Ref.~\onlinecite{Wang20}.
The total Lagrangian density is given by ${\cal L} = {\cal L}_f(- a_{\mu}) + {\cal L}_b(A_{\mu}, a_{\mu}, b_{\mu \nu}, c_{\mu})$, where 
${\cal L}_f(- a_{\mu})$ is given by Eq.~\eqref{eq:11} plus an FFLO (or strong enough BCS) pairing term, which gaps out the Weyl nodes. 
Spinon pairing produces a Meissner term for the gauge field $a_{\mu}$. Since ${\cal L}_f$ is defined on a lattice and $a_{\mu}$ is a compact 
gauge field, the Meissner term has the form $- \cos(2 a_{\mu})$, which makes $a_{\mu}$ a ${\mathbb Z}_2$ gauge field. 
A vison excitation of this ${\mathbb Z}_2$ gauge field corresponds to $\pi$ flux ($h c/ 2 e$ flux in normal units), which is the superconducting flux quantum.~\cite{Senthil_Z2}

Integrating out $b_{\mu \nu}$ in Eq.~\eqref{eq:24} produces a Meissner term for the combination $A_{\mu} + a_{\mu} + 2 c_{\mu}$, which means that at low energies 
we have
\beq
\label{eq:25}
c_{\mu} = - \frac{A_{\mu} + a_{\mu}}{2}.
\eeq
This makes $c_{\mu}$ a ${\mathbb Z}_4$ gauge field, since $c_{\mu}$ is also compact. 
This emergent ${\mathbb Z}_4$ gauge theory structure corresponds to the following electron fractionalization pattern
\beq
\label{eq:26}
\psi = f b_1 b_2, 
\eeq
where $b_{1,2}$ are charge-$1/2$ bosons.  We will discuss the physical meaning of this fractionalization in greater detail below. 

Let us now minimally couple the gauge fields $b_{\mu \nu}$ and $c_{\mu}$ to the corresponding conserved current sources $j_{\mu \nu}$ 
and $j_{\mu}$
\beqa
\label{eq:27}
{\cal L}_b&=&\frac{i}{2 \pi} (A_{\mu} + 2 c_{\mu}) \epsilon_{\mu \nu \lambda \rho} \partial_{\nu} b_{\lambda \rho} - \frac{2 i}{4 \pi} \epsilon_{z \mu \nu \lambda} 
c_{\mu} \partial_{\nu} c_{\lambda} \nonumber \\
&+&i b_{\mu \nu} j_{\mu \nu} + i c_{\mu} j_{\mu}, 
\eeqa
where we have ignored the coupling to the spinons at this point. 
The currents $j_{\mu \nu}$ and $j_{\mu}$ describe vortex loop and quasiparticle excitations correspondingly. 
Let us first set $j_{\mu \nu} = 0$ and integrate out $b_{\mu \nu}$. 
This gives
\beq
\label{eq:28}
{\cal L}_b = - \frac{i}{8 \pi} \epsilon_{z \mu \nu \lambda} A_{\mu} \partial_{\nu} A_{\lambda} - \frac{i}{2} j_{\mu} A_{\mu}. 
\eeq
The first term gives electrical Hall conductivity
\beq
\label{eq:29}
\sigma_{xy} = \frac{1}{4 \pi} = \frac{e^2}{h} \frac{\pi}{2 \pi}, 
\eeq
which is identical to the Hall conductivity of a Weyl semimetal with $Q = \pi/2$. 
The second term tells us that quasiparticle excitations are bosons that carry charge-$1/2$. 
\begin{figure}[t]
\hspace{-1cm}
\includegraphics[width=9cm]{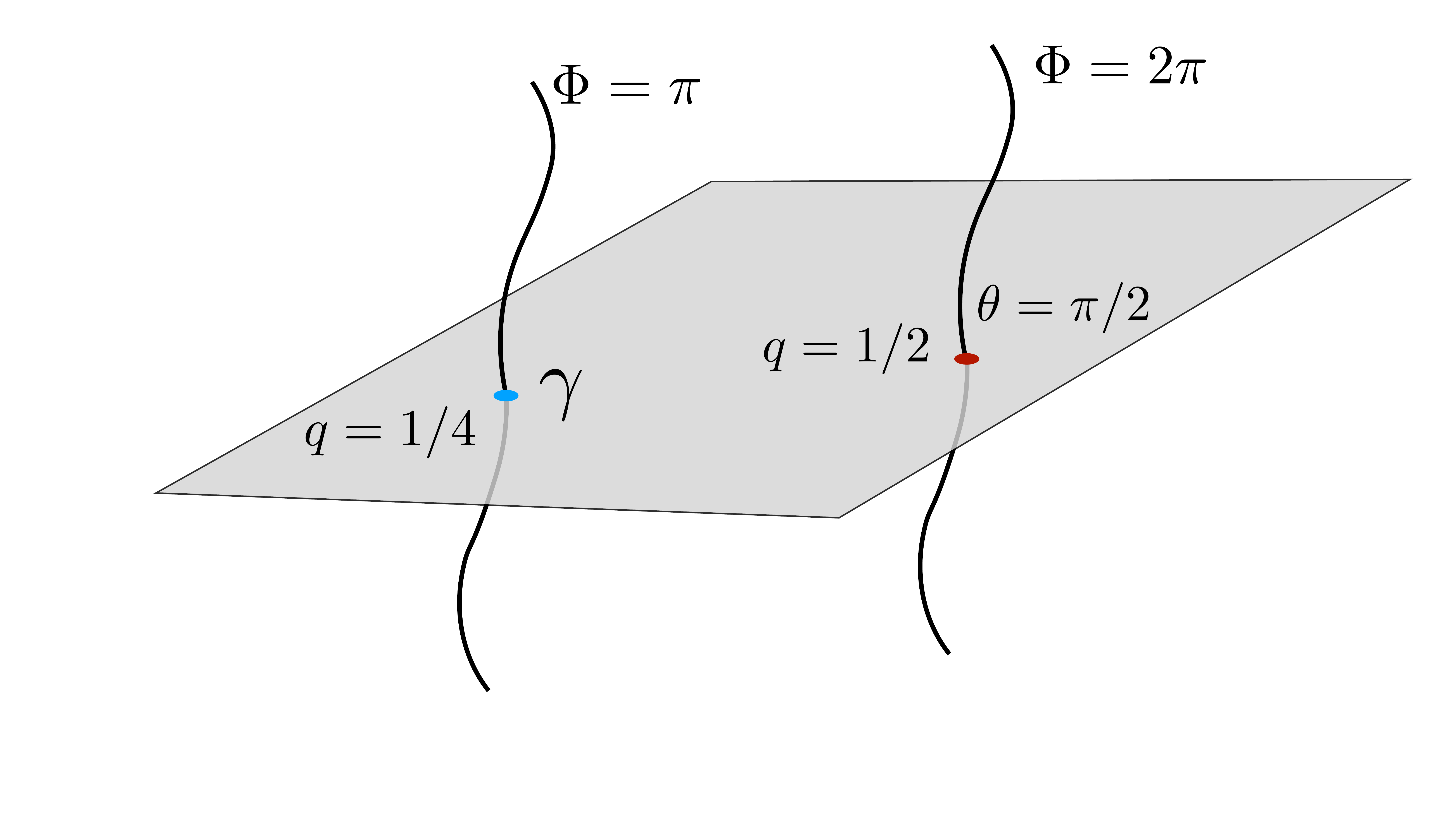}
\caption{(Color online) Induced charges and exchange statistics for intersections of flux $\pi$ and $2 \pi$ vortex lines with the $xy$-plane. 
A $\pi$-flux line induces a charge-$1/4$ and a localized Majorana mode, while a $2 \pi$-flux line induces a charge-$1/2$ semion.}
\label{fig:2}
\end{figure}

Now let us set $j_{\mu} = 0$ instead and integrate out $c_{\mu}$. 
We obtain
\beq
\label{eq:30}
\frac{\delta {\cal L}_b}{\delta c_{\mu}} = - \frac{i}{\pi} \epsilon_{z \mu \nu \lambda} \partial_{\nu} c_{\lambda} + \frac{i}{\pi} \epsilon_{\mu \nu \lambda \rho} \partial_{\nu}
b_{\lambda \rho} = 0. 
\eeq
We may solve this equation assuming that all fields are uniform in the $z$-direction. This corresponds to a mean-field picture of our 3D incompressible liquid as a stack of independent 
2D incompressible liquids. 
This gives
\beq
\label{eq:31}
c_{\mu} = - 2 b_{\mu z}. 
\eeq

What is the physical meaning of $b_{\mu z}$? Consider $j_{\mu z}$, which is the component of the vortex current, minimally coupled to $b_{\mu z}$. 
If all other components of $j_{\mu \nu}$ are zero, $j_{\mu z}$ corresponds to a straight-line vortex parallel to the  $z$-axis.
Intersection of this vortex line with the $xy$-plane may be viewed as a particle, to which the one-form gauge field $b_{\mu z}$ is minimally coupled. 
Plugging Eq.~\eqref{eq:31} into Eq.~\eqref{eq:27}, and rescaling variables $b_{\mu z} \rightarrow b_{\mu z}/2$ we obtain
\beq
\label{eq:32}
{\cal L}_b = \frac{2 i}{4 \pi} \epsilon_{z \mu \nu \lambda} b_{\mu z} \partial_{\nu} b_{\lambda z} - \frac{i}{2 \pi} A_{\mu} \epsilon_{z \mu \nu \lambda} \partial_{\nu} b_{\lambda z} 
+ i b_{\mu z} j_{\mu z}. 
\eeq
This describes a stack of 2D FQH liquids of bosons in the $\nu = 1/2$ Laughlin state. 
Integrating out $b_{\mu z}$ we obtain
\beq
\label{eq:33}
{\cal L}_b = - \frac{i}{8 \pi} \epsilon_{z \mu \nu \lambda} A_{\mu} \partial_{\nu} A_{\lambda} + \frac{i}{2} A_{\mu} j_{\mu z} - \frac{i}{8 \pi} \epsilon_{z \mu \nu \lambda} 
\tilde c_{\mu} \partial_{\nu} \tilde c_{\lambda}, 
\eeq
where we have written $j_{\mu z} = \epsilon_{z \mu \nu \lambda} \partial_{\nu} \tilde c_{\lambda}/2 \pi$ for convenience. 
Eq.~\eqref{eq:33} tells us that intersection of a vortex line, parallel to $z$, with the $xy$-plane, behaves as a particle of charge-$1/2$ and semionic 
statistics $\theta = \pi/2$, see Fig.~\ref{fig:2}. 
As discussed in Ref.~\onlinecite{Wang20}, this semionic self-statistics of an isolated intersection of a $2 \pi$ vortex with the $xy$-plane, which occurs, for example, when a dislocation 
with the Burgers vector along the $z$-axis is inserted into the system, is what prevents condensation of $2 \pi$ vortex loops. 
In our derivation in the previous section this was manifested in the impossibility of writing down a theory with $2 \pi$ vortices condensed, i.e. with a unit coefficient in 
front of $c_{\mu}$ in the first term in Eq.~\eqref{eq:24}, while maintaining a properly quantized coefficient of the Chern-Simons term in~\eqref{eq:24} and getting the right 
electromagnetic response Eq.~\eqref{eq:29}. 

We have so far established that the charge sector of our theory ${\cal L}_b$ correctly reproduces topological part of the electromagnetic response of the Weyl semimetal,
namely the electrical Hall conductivity, given by Eq.~\eqref{eq:29}. 
We want to also reproduce the thermal response, which in the noninteracting Weyl semimetal is tied to the electrical response by the Wiedemann-Franz law
\beq
\label{eq:34}
\kappa_{xy} = \sigma_{xy} \frac{\pi^2 k_B^2 T}{3}. 
\eeq
We note that the fermionic sector of the theory ${\cal L}_f$ already produces the right thermal Hall conductivity $\kappa_{xy}$, which arises from the chiral Majorana surface state, 
spanning the BZ. 
This means that ${\cal L}_b$ must describe a state with zero thermal Hall conductivity, if our 3D incompressible liquid indeed fully reproduces the chiral anomaly of a weakly-interacting Weyl semimetal. 

\begin{figure}[t]
\hspace{-0.5cm}
\includegraphics[width=9cm]{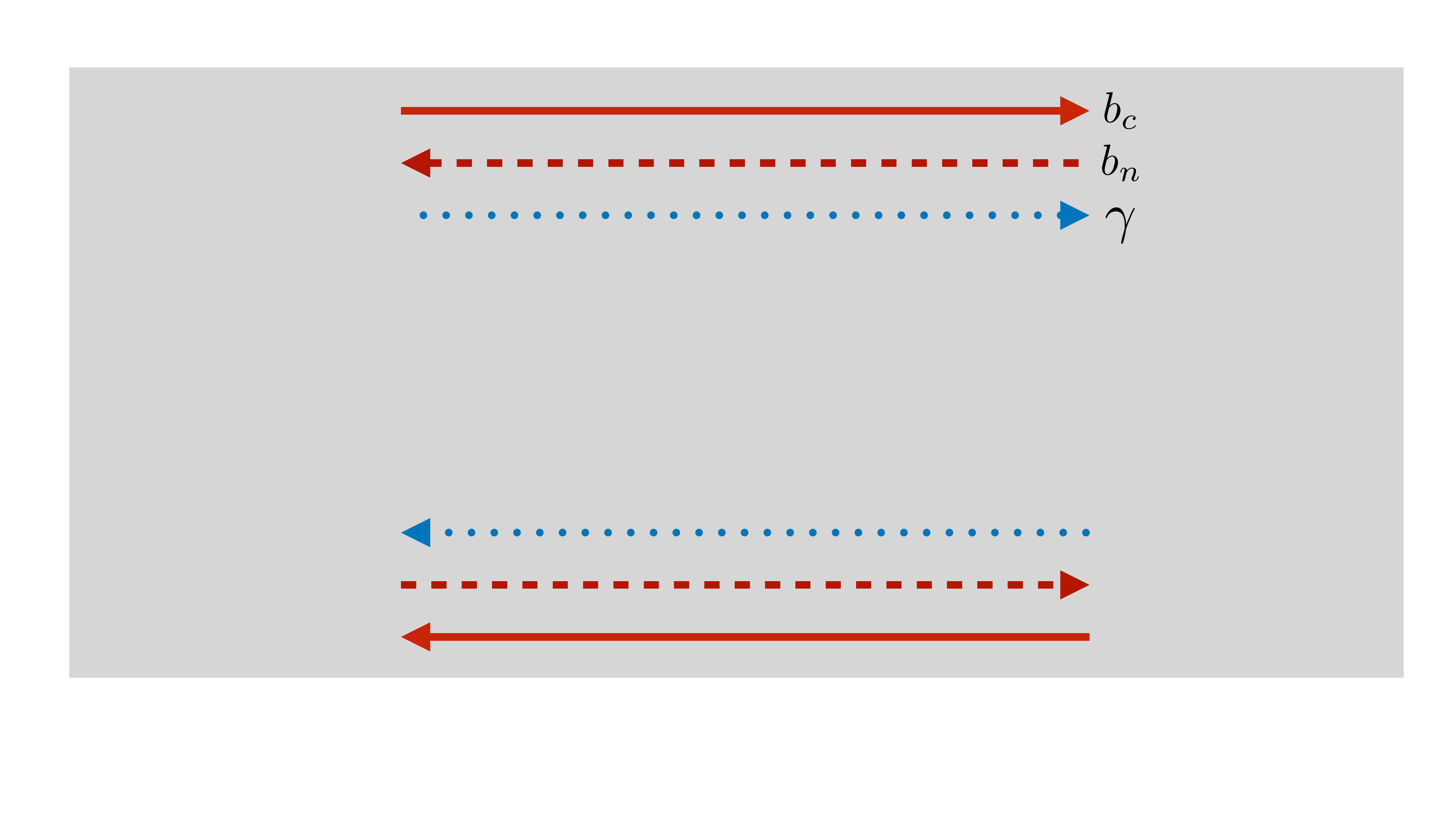}
\caption{(Color online) Schematics of the surface state structure of the 3D FQH liquid. Only a 2D section, perpendicular to the $z$-axis, is shown for simplicity. 
There are two opposite-chirality bosonic modes: charged (red solid line, $b_c$) and neutral (red dashed line, $b_n$), and a Majorana mode (blue dotted line, $\gamma$).}
\label{fig:3}
\end{figure}

The simplest way to see that this is indeed the case is to invoke the mean-field approximation, described above, in which only $b_{\mu z}$ components of the two-form gauge field $b_{\mu \nu}$ are nonzero. 
Then we obtain
\beq
\label{eq:35}
{\cal L}_b = - \frac{i}{\pi} (A_{\mu} + 2 c_{\mu}) \epsilon_{z \mu \nu \lambda} \partial_{\nu} b_{\lambda z} - \frac{i}{2 \pi} \epsilon_{z \mu \nu \lambda} 
c_{\mu} \partial_{\nu} c_{\lambda}. 
\eeq
Changing variables as $b_{\mu z} \rightarrow (b_{\mu z} - c_{\mu})/4$, this becomes
\beqa
\label{eq:36}
{\cal L}_b&=&- \frac{i}{4 \pi} \epsilon_{z \mu \nu \lambda} c_{\mu} \partial_{\nu} b_{\lambda z} - \frac{i}{4 \pi} \epsilon_{z \mu \nu \lambda} b_{\mu z} \partial_{\nu} c_{\lambda} 
\nonumber \\
&-&\frac{i}{4 \pi} \epsilon_{z \mu \nu \lambda} A_{\mu} \partial_{\nu} (b_{\lambda z} - c_{\lambda}). 
\eeqa
This describes a stack of ``bosonic integer quantum Hall" states~\cite{Lu-Vishwanath,Senthil-Levin} of two-component charge-$1/2$ bosons, which 
may be viewed as the $b_{1,2}$ in Eq.~\eqref{eq:26}. 
As can be easily seen by diagonalizing the $K$-matrix, corresponding to Eq.~\eqref{eq:36}, i.e. $K = \sigma_x$, this theory 
contains two edge modes: one charged, which gives the Hall conductivity of Eq.~\eqref{eq:29}, and one neutral, which has opposite chirality, see Fig.~\ref{fig:3}. 
Thus the thermal Hall conductivity in this state is indeed zero and we have a gapped incompressible liquid state, which has the same topological 
response (chiral anomaly) as a noninteracting Weyl semimetal. 

One final issue we have not yet touched upon is excitations of this incompressible fractionalized 3D liquid that arise from the fermionic sector of the theory. 
Quasiparticle fermionic excitations are the neutral spinons $f$ themselves. 
There are also vison loop excitations of the gauge field $a_{\mu}$. To understand their properties let us go back to Eq.~\eqref{eq:25}. 
Substituting this into the full Lagrangian, we obtain
\beqa
\label{eq:37}
{\cal L}&=&{\cal L}_f(- a_{\mu}) - \frac{i}{8 \pi} \epsilon_{z \mu \nu \lambda} A_{\mu} \partial_{\nu} A_{\lambda} \nonumber \\
&-&\frac{i}{4 \pi} \epsilon_{z \mu \nu \lambda} A_{\mu} \partial_{\nu} a_{\lambda} - \frac{i}{8 \pi} \epsilon_{z \mu \nu \lambda} a_{\mu} \partial_{\nu} a_{\lambda}.
\eeqa
Consider a straight vison (i.e. $\pi$-flux of $a_{\mu}$) line, parallel to the $z$-axis. 
The above equation tells us that such a vortex line induces a charge-$1/4$ when intersecting an $xy$-plane. 
In addition, such a vortex binds a 1D helical Majorana mode, dispersing along the $z$-direction,
which may be easily obtained by solving the corresponding Bogoliubov-de Gennes equation.~\cite{Wang20}
It follows that an intersection of the vison loop with an atomic $xy$-plane induces a $1/4$ charge and a localized zero-energy Majorana mode, as illustrated in Fig.~\ref{fig:2}.
Braiding of such vison loop excitations, when linked with an isolated dislocation line, is characterized by nonabelian statistics due to the presence of 
the Majorana mode. 
The 3D FQH state we have found may thus be viewed as a 3D analog of nonabelian even-denominator 2D FQH liquids. 
\section{Conclusions}
\label{sec:4}
In this paper we have discussed hydrodynamic theory of the 3D FQH state, obtained after gapping Weyl nodes in a magnetic Weyl semimetal without breaking translational 
symmetry, introduced in Ref.~\onlinecite{Wang20}. 
This takes the form of a hydrodynamic BF theory, which is a 3D analog of the Chern-Simons theory of 2D FQHE. 
An important difference from the 2D case is that this theory contains two gauge fields: a two-form gauge field $b_{\mu \nu}$ and a one-form gauge field $c_{\mu}$. 
Physically, this expresses the existence of two kinds of excitations in the 3D FQH liquid: quasiparticle excitations, which couple to $c_{\mu}$, and loop excitations, which couple to $b_{\mu \nu}$. 
In addition, there is a statistical gauge field $a_{\mu}$, which couples bosonic and fermionic sectors of the theory. 
Unlike in 2D FQHE, quasiparticle excitations are always either bosons or fermions, as there is no fractional statistics in 3D. 
The closest analog of the fractionally-charged anyon excitations of 2D FQH liquids are in fact the loop excitations. In particular, when a $2 \pi$ vortex loop 
intersects an $xy$ atomic plane, the intersection point carries charge-$1/2$. Exchange of two such intersection points may be sharply defined when a pair of vortex loops 
is linked with a dislocation line in the $xy$-plane.~\cite{Wang-Levin,Wang20} In this case, the exchange statistics of the linked $2 \pi$ vortex loops is semionic. 
A $\pi$-flux vortex loop induces a charge-$1/4$ and a localized Majorana mode, leading to nonabelian exchange statistics. 

While in this paper and in Ref.~\onlinecite{Wang20} we focused on fully gapped states, preserving the chiral anomaly of a noninteracting Weyl semimetal, it also makes sense 
to consider gapless strongly correlated states with the same property. 
Such states may be accessed easily within the formalism, presented above, by simply leaving the spinons unpaired, while still placing the chargons in a gapped fractional 
quantum Hall state. This type of gapless topologically-ordered states may be regarded as 3D analogs of the composite fermion Fermi liquid at filling factor $\nu = 1/2$,~\cite{HLR}
and may be called ``composite Weyl liquids". While a similar term was introduced earlier in Ref.~\onlinecite{Sagi18}, its meaning is somewhat different in our case. 
Unlike the fully gapped 3D FQH liquid, nontrivial composite Weyl liquids may presumably exist at many different values of the Weyl node separation. 
The corresponding topological orders may be described by condensing vortex composites with even integer vorticity larger than four in units of the superconducting flux quantum. 
We leave a more detailed discussion of possible composite Weyl liquids and their physical properties to future work. 

\begin{acknowledgments}
We thank Yuan-Ming Lu and Chong Wang for useful discussions. 
MT was supported by the Natural Sciences and Engineering Research Council (NSERC) of Canada.
AAB was supported by Center for Advancement of Topological Semimetals, an Energy Frontier Research Center funded by the U.S. Department of Energy Office of Science, Office of Basic Energy Sciences, through the Ames Laboratory under
contract DE-AC02-07CH11358. 
AAB also acknowledges the hospitality of KITP, where part of this work was performed. Research at KITP is supported in part by the National Science Foundation under Grant No. NSF PHY-1748958.
\end{acknowledgments}
\bibliography{references}
\end{document}